\input ./style/arxiv-ba.cfg
\documentclass[ba,preprint,linksfromyear]{imsart}
\makeatletter
   \@ifpackageloaded{natbib}{}{\usepackage{natbib}}
\makeatother

\pubyear{2015}
\volume{10}
\issue{3}
\firstpage{605}
\lastpage{625}
\doi{10.1214/14-BA924}

\startlocaldefs
\def\bysame{---}
\setattribute{email}{text}{}
\newcommand{\enquote}[1]{``#1''}
\expandafter\ifx\csname natexlab\endcsname\relax\def\natexlab#1{#1}\fi

\endlocaldefs
\begin{document}

\begin{frontmatter}
\title{Bayesian Nonparametric Weighted Sampling Inference}
\runtitle{Bayesian Weighted Inference}

\begin{aug}
\author[a]{\fnms{Yajuan} \snm{Si}\corref{}\ead[label=e1]{ysi@biostat.wisc.edu}},
\author[b]{\fnms{Natesh S.} \snm{Pillai}\ead[label=e2]{pillai@fas.harvard.edu}}, and
\author[c]{\fnms{Andrew} \snm{Gelman}\ead[label=e3]{gelman@stat.columbia.edu}}

\address[a]{Department of Biostatistics \& Medical Informatics and Department of
Population Health Sciences, University of Wisconsin, Madison, WI, \printead{e1}}
\address[b]{Department of Statistics, Harvard University, Cambridge,
MA, \printead{e2}}
\address[c]{Department of Statistics and Department of Political Science, Columbia University,
New York, NY, \printead{e3}}
\end{aug}
\runauthor{Yajuan Si, Natesh S. Pillai, and Andrew Gelman}

\begin{abstract}
It has historically been a challenge to perform Bayesian inference in a
design-based survey context. The present paper develops a Bayesian
model for sampling inference in the presence of inverse-probability
weights. We use a hierarchical approach in which we model the
distribution of the weights of the nonsampled units in the population
and simultaneously include them as predictors in a nonparametric
Gaussian process regression. We use simulation studies to evaluate the
performance of our procedure and compare it to the classical
design-based estimator. We apply our method to the Fragile Family and Child
Wellbeing Study. Our studies find the Bayesian nonparametric finite
population estimator to be more robust than the classical design-based
estimator without loss in efficiency, which works because we induce
regularization for small cells and thus this is a way of automatically
smoothing the highly variable weights.
\end{abstract}



\end{frontmatter}


\section{Introduction}
 \label{introduction}

\subsection{The problem}

    Survey weighting adjusts for known or expected differences between sample and population. These differences arise from sampling design, undercoverage, nonresponse, and limited sample size in subpopulations of interest. Weights are constructed based on design or benchmarking calibration variables that are predictors of inclusion probability $\pi_i$, defined as the probability that unit $i$ will be included in the sample, where inclusion refers to both selection into the sample and response given selection. However, weights have problems. As pointed out by \citet{gelman07} and the associated discussions, current approaches for construction of weights are not systematic, with much judgment required on which variables to include in weighting, which interactions to consider, and whether and how weights should be trimmed.

    In applied survey research it is common to analyze survey data collected by others in which weights are included, along with instructions such as  ``This file contains a weight variable that must be used in any analysis'' \citep{cbs88} but without complete information on the sampling and weighting process. In classical analysis, externally-supplied weights are typically taken to be inversely proportional to each unit's probability of inclusion into the sample. Current Bayesian methods typically ignore survey weights, assume that all weighting information is included in regression adjustment, or make \emph{ad hoc} adjustments to account for the design effect induced by unequal weights \citep{ghitza:gelman:13}.  A fully Bayesian method accounting for survey weights would address a serious existing concern with these methods.

    Our goal here is to incorporate externally-supplied sampling weights in Bayesian analysis, continuing to make the assumption that these weights represent inverse inclusion probabilities but in a fully model-based framework. This work has both theoretical and practical motivations. From a theoretical direction, we would like to develop a fully nonparametric version of model-based survey inference. In practice, we would like to get the many benefits of multilevel regression and poststratification (see, \emph{e.g.}, \citealt{lax:philips:09a,lax:philips:09b}) in realistic settings with survey weights. The key challenge is that, in a classical survey weighting setting, the population distribution of the weights is not known, hence the poststratification weights must themselves be modeled.

    In this paper we consider a simple survey design using independent sampling with unequal inclusion probabilities for different units and develop an approach for fully Bayesian inference using externally-supplied survey weights. This framework is limited (for example, it does not allow one to adjust for cluster sampling) but it is standard in applied survey work, where users are supplied with a publicly released dataset and sampling weights but little or no other design information. In using only the weights, we seek to generalize classical design-based inference, more fully capturing inferential uncertainty and in a more open-ended framework that should allow more structured modeling as desired. For simplicity, we consider a univariate survey response, with the externally-supplied weight being the only other information available about the sample units. Our approach can be generalized to multivariate outcomes and to include additional informative predictors, which will be elaborated in Section~\ref{conclusion}.

\subsection{Background}

    Consider a survey response with population values $Y_i$, $i=1,\dots, N$, with the sampled values labeled $y_i$, for $i=1,\dots,n$, where $N$ is the population size and $n$ is the sample size. Design-based inference treats $Y_i$ in the population as fixed and the inclusion indicators $I_i$ as random. Model-based survey inference builds models on $y_i$ and uses the model to predict the nonsampled values $Y_i$. \citet{rubin83-pi} and \citet{little83-pi} point out that, to satisfy the likelihood principle, any model for survey outcomes should be conditional on all information that predicts the inclusion probabilities. If these probabilities take on discrete values, this corresponds to modeling within strata defined by the inclusion probabilities.

    We assume that the survey comes with weights that are proportional to the inverse of the inclusion probabilities, so that we can map the unique values of sample weights to form the strata. We call the strata {\em poststratification cells} since they are constructed based on the weights in the sample. This definition is different from the classical poststratification (also called benchmarking calibration) adjustment, which is an intermediate step during weight construction.  In the present paper we assume we are given inverse-probability weights without any further information on how they arose.

    We use the unified notation for weighting and poststratification from \citet{little91,little93}. Suppose the units in the population and the sample can be divided into $J$ poststratification cells \citep{gelman:little:97,gelmancarlin01} with population cell size $N_j$ and sample cell size $n_j$ for each cell $j=1,\dots, J$, with $N=\sum_{j=1}^JN_j$ and $n=\sum_{j=1}^Jn_j$.  For example, in a national political survey, the cells might be a cross-tabulation of 4 age categories, 2 sexes, 4 ethnicity categories, 5 education categories and 50 states, so that the resulting number of cells $J$ is $4\times2\times4\times5\times50$. The sample sizes $n_j$'s can be small or even zero. In the present paper, however, we shall define poststratification cells based on the survey weights in the observed data, where the collected $n_j \geq 1$.  Thus, even if the cells are classified by sex, ethnicity, \emph{etc.}, here we shall just treat them as $J$ cells characterized only by their weights.

    As in classical sampling theory, we shall focus on estimating the population mean of a single survey response: $\theta=\overline{Y}=\frac{1}{N}\sum_{i=1}^NY_i$.  Let $\theta_j=\overline{Y}_j$ be the population mean within cell $j$, and let $\bar{y}_j$ be the sample mean. The overall mean in the population can be written as
$$
    \theta=\overline{Y}=\sum_{j=1}^J\frac{N_j}{N}\theta_j.
$$

    Classical design-based approaches use the unsmoothed estimates $\tilde{\theta}_j=\bar{y}_j$ without any modeling of the survey response, and the cell allocation depends on the $n_j$'s and $N_j$'s. The population mean estimate is expressed as
    \begin{equation}
    \label{design}
    \tilde{\theta}^{\,\textrm{design-based}}=\sum_{j=1}^J \frac{N_j}{N}\bar{y}_j=\sum_{i=1}^n \frac{N_{j[i]}}{n_{j[i]}N}y_i,
    \end{equation}
where $j[i]$ is the poststratification cell containing individual $i$, and $\frac{N_{j[i]}}{n_{j[i]}N}$ defined as equal to $w_{j[i]}$ is the unit weight of any individual $i$ within cell $j$. This design-based estimate is unbiased. However, in realistic settings, unit weights $w_{j}$'s result from various adjustments. With either the inverse-probability unit weights $w_j \propto \frac{1}{\pi_j}$ or the benchmarking calibration unit weights $w_j\propto \frac{N_j^*}{n_j^*}$, the classical weighted ratio estimator \citep{ht52,hajek71} is
    \begin{equation}
    \label{ht}
    \tilde{\theta}^{\,\mathrm{design}}=\frac{\sum_{i=1}^nw_{j[i]}y_i}{\sum_{i=1}^n w_{j[i]}}.
    \end{equation}
Note that $N_j^*$ and $n_j^*$ are population and sample size for cell $j$ defined by the calibration factor, respectively. This cell structure depends on the application study. We use $*$ to distinguish them from the $N_j$'s and $n_j$'s defined in the paper. In classical weighting with many factors, the unit weights $w_j$ become highly variable, and then the resulting $\tilde{\theta}$ becomes unstable. In practice the solution is often to trim the weights or restrict the number of adjustment variables, but then there is a concern that the weighted sample no longer matches the population, especially for subdomains.

    Our proposed estimate for $\theta$ can be expressed in weighted form,
    \begin{equation}
    \label{model-based}
    \tilde{\theta}^{\,\mathrm{model-based}}=\sum_{j=1}^J\frac{N_j}{N}\tilde{\theta}_j.
    \end{equation}
    This provides a unified expression for the existing estimation procedures. The $N_j/N$'s represent the allocation proportions to different cells. We will use modeling to obtain the estimates $\tilde{\theta}_j$ from the sample means $\bar{y}_j$, sample sizes $n_j$ and unit weights $w_j$. Inside each cell the unit weights are equal, by construction. Design-based and model-based methods both implicitly assume equal probability sampling within cells. Thus $\bar{y}_j$ can be a reasonable estimate for $\theta_j$, where the model-based estimator is equivalent to the design-based estimator defined in (\ref{design}). When cell sizes are small, though, multilevel regression can be used to obtain better estimates of $\theta_j$'s.

    Model-based estimates are subject to bias under misspecified model assumptions. Survey sampling practitioners tend to avoid model specifications and advocate design-based methods. Model-based methods should be able to yield similar results as classical methods when classical methods are reasonable. This means that any model for survey responses should at least include all the information used in weighting-based methods. In summary, model-based inference is a general approach but can be difficult to implement if the relevant variables are not fully observed in the population. Design-based inference is appealingly simple, but there are concerns about capturing inferential uncertainty, obtaining good estimates in subpopulations, and adjusting for large numbers of design factors.

    If full information is available on design and nonresponse models, the design will be ignorable given this information. This information can be included as predictors in a regression framework, using hierarchical modeling as appropriate to handle discrete predictors (for example, group indicators for clusters or small geographic areas, such as the work by \citealt{oleson07}) to borrow strength in the context of sparse data. However, it is common in real-world survey analysis for the details of sampling and nonresponse not to be supplied to the user, who is given only unit-level weights and perhaps some information on primary sampling units. The methods we develop here are relevant for that situation, in which we would like to enable researchers to perform model-based and design-consistent Bayesian inference using externally-supplied sample survey weights.

 \subsection{Contributions of this paper}

    Existing methods for the Bayesian analysis of weighted survey data, with the exception of \citet{little12}, assume the weights to be known for all population units, but we assume that the weights are known for the sample units only, which is more common in practice. When only sample records are available in probability-proportional-to-size selection, \citet{little12} implement constrained Bayesian bootstrapping for estimating the sizes of nonsampled units and screen out the estimates whose summations are the closest to the true summation of sizes. The probabilities for this Bayesian bootstrap depend only on the available sample sizes and do not account for the design. The estimated sizes are used as covariates in a regression model for predicting the outcomes. Even though multiple draws are kept for the measure sizes to capture the uncertainty, the two steps are not integrated.

    In contrast, we account for the unequal probabilities of inclusion and construct a joint model for weights and outcomes under a fully Bayesian framework. Conceptually, this is a key difference between our work and previously published methods. Furthermore, our approach is applicable to most common sampling designs, whereas \citet{little12} assume probability-proportional-to-size sampling and do not consider the selection probability when predicting non-sampled sizes. Elsewhere in the literature, \cite{elliott00} and \cite{elliott07} consider discretization of the weights and assume that unique values of the weights determine the poststratification cells. But these authors assume the weights are known for the population and thus do not estimate the
    weights.\looseness=-1

    We can now use our conceptual framework for jointly modeling both the weights and the outcomes, to incorporate nonparametric approaches for function estimation. We use the log-transformed weights as covariates, which makes sense because the weights in survey sampling are often obtained by multiplying different adjustment factors corresponding to different stages of sampling, nonresponse, poststratification, \emph{etc}.  Our proposed approach takes advantage of the well known theoretical properties and computational advantages of Gaussian processes, but of course other methods may be used as well. The idea of nonparametric modeling for the outcomes given the weights is a natural extension of work by previous authors \citep[\emph{e.g.},][]{elliott00, zhenglittle03, elliott07, qchen10}.

    We call our procedure Bayesian nonparametric finite population (BNFP) estimation. We fit the model using the Bayesian inference package Stan \citep{stan-software:2013,stan-manual:2013}, which obtains samples from the posterior distribution using the no-U-turn sampler, a variant of Hamiltonian Monte Carlo \citep{hoffman-gelman:2012}.  In addition to computational efficiency, being able to express the model in Stan makes this method more accessible to practitioners and also allows the model to be directly altered and extended.


\section{Bayesian model for survey data with externally-supplied weights}
\label{methods}

    In the collected dataset, we observe $(w_{j[i]}, y_i)$, for $i=1,\dots,n$, with the weight $w_j$ being proportional to the inverse probability of sampling for units in cell $j$, for $j=1,\dots,J$, and unique values of weights determine the $J$ cells. We assume that the values of the weights in the sample represent all the information we have about the sampling and weighting procedure.

    In our Bayesian hierarchical model, we simultaneously predict the $w_{j[i]}$'s and $y_i$'s for the $N-n$ nonsampled units. First, the discrete distribution of $w_{j[i]}$'s is determined by the population cell sizes $N_j$'s, which are unknown and treated as parameters in the model for the sample cell counts $n_j$'s. Second, we make inference based on the regression model $y_i|w_{j[i]}$ in the population under a Gaussian process (GP) prior for the mean function; for example, when $y_i$ is continuous: $y_i|w_{j[i]} \sim \mbox{N}(\mu(\log w_{j[i]}), \sigma^2)$ with prior distribution $\mu(\cdot)\sim \mathrm{GP}$. Finally, the two integrated steps generate the predicted $y_i$'s in the population. Our procedure is able to yield the posterior samples of the population quantities of interest; for instance, we use the posterior mean as the proposed estimate $\tilde{\theta}$. We assume the population size $N$ is large enough to ignore the finite population correction factors. We describe the procedure in detail in the remainder of this section.

\subsection{Model for the weights}\label{bb}
\vspace*{-3pt}

    Our model assumes the unique values of the unit weights determine the $J$ poststratification cells for the individuals in the population via a one to one mapping. This assumption is not generally correct (for example, if weights are constructed by multiplying factors for different demographic variables, there may be some empty cells in the sample corresponding to unique products of factors that would appear in the population but not in the sample) but it allows us to proceed without additional knowledge of the process by which the weights were constructed.

    We assume independent sampling for the sample inclusion indicator with probability
$$
    \mbox{Pr}(I_{i}=1)=\pi_{i}=c/w_{j[i]}, \mbox{ for } i=1,\dots,N,
$$
     where $c$ is a positive normalizing constant, induced to match the probability of sampling $n$ units from the population. By definition, $I_{i} = 1$ for observed samples, \emph{i.e.}, $i = 1, \dots, n$.     We label $N_j$ as the number of individuals in cell $j$ in the population, each with weight $w_{j}$, and we have observed $n_j$ of them. The unit weight $w_{j}$ is equal to $w_{j[i]}$ for unit $i$ belonging to cell $j$ and its frequency in the population is $N_j$. Because units belonging to cell $j$ have equal inclusion probability $c/w_j$, the expectation for sample cell size $n_j$ is $\mbox{E}(n_j)=cN_j/w_j$. Surveys are typically designed with some pre-determined sample size in mind (even though it may not be true in practice where nonresponse contaminates), thus we treat $n$ as fixed. By $n=\sum_{j=1}^Jn_j$,  this implies $c=n\frac{1}{\sum_{j=1}^J N_j/w_{j}}$. We assume $n_j$'s follow a multinomial distribution conditional on $n$,
$$
 \vec{n}=(n_1,\dots, n_J)\sim \textrm{Multinomial}\left(n; \frac{N_1/w_{1}}{\sum_{j=1}^J N_j/w_{j}},\dots, \frac{N_J/w_{J}}{\sum_{j=1}^J N_j/w_{j}}\right).
$$

For each cell $j$, we need to predict the survey outcome for the nonsampled $N_j - n_j$ units with weight $w_{j}$. Here $N_j$'s are unknown parameters.  Since we assume there are only $J$ unique values of weights in the population, by inferring $N_j$, we are implicitly predicting the weights for all the nonsampled units.

\subsection{Nonparametric regression model for $y|w$}
\label{gp}
\vspace*{-3pt}

    Let $x_j=\log w_j$.  For a continuous survey response $y$, we work with the default model
$$
    y_i \sim \mbox{N}(\mu(x_{j[i]}), \sigma^2),
$$
    where $\mu(x_j)$ is a mean function of $x_j$ that we model nonparametrically. We use the logarithm of survey weights because such weights are typically constructed multiplicatively (as the product of inverse selection probabilities, benchmarking calibration ratio factors, and inverse response propensity scores), so that changes in the weights correspond to additive differences on the log scale.  We can re-express the model using the means and standard deviations of the response within cells as
$$
        \bar{y}_{j} \sim \mbox{N}(\mu(x_{j}),\sigma^2/n_j)\mbox{  and }
        \sum_{j=1}^Js_j^2/\sigma^2 \sim \chi^2_{n-1},
$$
based on the poststratification cell mean $\bar{y}_{j}=\sum_{i\,\in\textrm{cell}\, j}y_i/n_j$ and the cell total variance $s_j^2=\sum_{i\,\in\, \textrm{cell}\, j}(y_{i}-\bar{y}_j)^2 $, for $j=1,\dots,J$.
Alternatively, we can allow the error variance to vary by cell:
$$
    \bar{y}_{j} \sim \mbox{N}(\mu(x_{j}),\sigma_j^2/n_j)\mbox{  and }
        s_j^2/\sigma_j^2 \sim \chi^2_{n_j-1}.
$$
    If the data $y$ are binary-valued, we denote the sample total in each cell $j$ as $y_{(j)}=\sum_{i\,\in\, \textrm{cell}\, j}y_i$ and assume
 \begin{equation}\label{eqn:binary}
        y_{(j)}\sim \textrm{Binomial}(n_j,\mbox{logit}^{-1}(\mu(x_j)).
 \end{equation}
It is not necessary to use an over-dispersed model here because the function $\mu(\cdot)$ includes a variance component for each individual cell, as we shall discuss below.

    Penalized splines have been used to construct nonparametric random slope models in survey inference \citep[\emph{e.g.},][]{zhenglittle03}. However, in our experience, highly variable weights bring in difficulties for the selection and fitting of spline models. We avoid parametric assumptions or specific functional forms for the mean function and instead use a GP prior on $\mu(\cdot)$. GP models constitute a flexible class of nonparametric models that avoid the need to explicitly specify basis functions \citep{gp-rasmussen06}. The class of GPs has a large support on the space of smooth functions and thus has attractive theoretical properties such as optimal posterior convergence rates for function estimation problems \citep{gprate08}. They are also computationally attractive because of their conjugacy properties. GPs are further closely related to basis expansion methods and include as special cases various well-known models, including spline models. In this regard, our model supersedes those spline models that were earlier used in research on survey analysis.

    We assume
$$
        \mu(x_{j})\sim \mathrm{GP}(x_{j}\beta, C(\tau,l,\delta)),
$$
where by default we have set the mean function equal to a linear regression, $x_j\beta$, with $\beta$ denoting an unknown
coefficient. Here $C(\tau, l, \delta)$ denotes the covariance function, for which we use a squared exponential kernel: for any $x_j, x_{j'}$,
 \begin{equation}\label{eqn:cover}
 \mathrm{Cov}(\mu(x_j),\mu(x_{j'}))=\tau^2(1-\delta) e^{-\frac{(x_{j}-x_{j'})^2}{l^2}} + \tau^2\,\delta \,\mathrm{I}_{j=j'},
 \end{equation}
 where $\tau$, $l$ and $\delta$ $(\in [0,1])$ are unknown hyperparameters with positive values. The length scale $l$ controls the local smoothness of the sample paths of $\mu(\cdot)$. Smoother sample paths imply more borrowing of information from neighboring $w$ values. The nugget term $\tau^2\delta$ accounts for unexplained variability between cells. As special cases, with $\delta=0$ this reduces to the common GP covariance kernel with $\tau^2$ as the partial sill and $l$ as the length scale but without the nugget term; with $\delta=1$ this is a penalized Bayesian hierarchical model with independent prior distributions $\mbox{N}(x_{j}\beta, \tau^2)$ for each $\mu(x_{j})$. The vector $\vec{\mu} = (\mu(x_1), \dots, \mu(x_J))$ has a $J$-variate Gaussian distribution
  $$
  \vec{\mu} \sim \mathrm{N}(\vec{x}\beta,\Sigma),
  $$
   where $\vec{x}=(x_1,\dots, x_J)$ and $\Sigma_{jj'}= \mathrm{Cov}(\mu(x_j), \mu(x_{j'}))$ as given by (\ref{eqn:cover}).

\subsection{Posterior inference for the finite population}\label{post}

    Let $\vec{y}$ be the data vector. The parameters of interest are
$\phi=(\sigma, \beta,l,\tau, \delta, \vec{N})$, where $\vec{N}$ is the vector of cell sizes $(N_1,\dots,N_J)$.
The likelihood function for the collected data is
$$
p(\vec{y}|\phi,w)\propto p(\vec{y}|\mu, \vec{n},\sigma^2)p(\mu|w, \beta,l,\tau,\delta)p(\vec{n}|w,\vec{N}).
$$
The joint posterior distribution is
$$
    p(\phi|\vec{y},w)\propto p(\vec{y}|\mu, \vec{n},\sigma)p(\mu|w, \beta,l,\tau,\delta)p(\vec{n}|w,\vec{N})p(\phi).
$$

    Often there is no prior information for the weights, and to reflect this we have chosen prior distributions for $(\beta, \sigma, \tau, l)$ to be weakly informative but proper. We assume the following prior distributions:
    \begin{eqnarray}
\nonumber    \beta&\sim &\mathrm{N}(0, 2.5^2)\\
\nonumber    \sigma &\sim& \textrm{Cauchy}^{+}(0, 2.5\cdot\textrm{sd}(y))\\
\nonumber    \tau &\sim &\textrm{Cauchy}^{+}(0, 2.5)\\
\nonumber    l &\sim &\textrm{Cauchy}^{+}(0, 2.5)\\
\nonumber    \delta &\sim& \textrm{U}[0,1]\\
    \pi(N_j)&\propto &1 \mbox{, for } j=1,\dots, J.
    \label{stan-prior-2}
    \end{eqnarray}

Here Cauchy$^{+}(a,b)$ denotes the positive part of a Cauchy distribution with center $a$ and scale $b$, and $\textrm{sd}(y)$ is the standard deviation of $y$ in the sample. Following the principle of scaling a weakly-informative prior with the scale of the data in \cite{weaklprior:gelman:08}, we assign the half-Cauchy prior with scale $2.5\cdot\textrm{sd}(y)$ to the unit-level scale $\sigma$. This is approximate because the prior is set from the data and is weakly informative because $2.5$ is a large factor and the scaling is based on the standard deviation of the data---not of the residuals---but this fits our goal of utilizing weak prior information to contain the inference. In any given application, stronger prior information could be used (indeed, this is the case for the other distributions in our model, and for statistical models in general) but our goal here is to define a default procedure, which is a long-respected goal in Bayesian inference and in statistics more generally. Furthermore, the weakly informative half-Cauchy prior distributions for the group-level scale parameter $\tau$ and smoothness parameter $l$ allow for values close to $0$ and heavy tails \citep{gelman06-prior,polson:scott:12}.

We are using uniform prior distributions on the allocation probability $\delta$ ($\in[0,1]$) and on the population poststratification cell sizes $N_j$'s (with nonnegative values). The population estimator will be normalized by the summation of the $N_j$'s to match the population total $N$. The weights themselves determine the poststratification cell structure, and the $N_j$'s represent the discrete distribution of the weights $w_j$'s in the population. But we do not have extra information and can only assign flat prior distributions. If we have extra information available on the $N_j$'s, we can include this into the prior distribution. For example, if some variables used to construct weights are available, we can incorporate their population frequencies as covariates to predict the $N_j$'s.

    Our objective is to make inference for the underlying population. For a continuous outcome variable, the population mean is defined as
$$
    \overline{Y} = \frac{\sum_{j=1}^J\overline{Y}_jN_j}{\sum_{j=1}^J N_j} = \frac{1}{\sum_{j=1}^J N_j}\sum_{j=1}^J(\bar{y}_j  n_j+ \bar{y}_{\mathrm{exc}, j} (N_j-n_j)),
$$
where $\bar{y}_{\mathrm{exc}, j}$ is the mean for nonsampled units in cell $j$, for $j=1,\dots, J$. The posterior predictive distribution for $\bar{y}_{\mathrm{exc}, j}$ is
$$
    (\bar{y}_{\mathrm{exc}, j}|-) \sim \mbox{N}(\mu(x_{j}),\sigma^2/(N_j-n_j)).
$$
When the cell sizes $N_j$'s are  large enough, $\bar{y}_{\mathrm{exc},j}$ is well approximated by $\mu(x_j)$.

    For a binary outcome, we are interested in the proportion of \emph{Yes} responses in the population:
$$
    \overline{Y}=\frac{\sum_{j=1}^J Y_{(j)}}{\sum_{j=1}^J N_j} = \frac{1}{\sum_{j=1}^J N_j}\sum_j\left(y_{(j)} + \sum_{i=1}^{N_j-n_j}y_{i}\right) =\frac{1}{\sum_{j=1}^J N_j}\sum_{j=1}^J\left(y_{(j)} + y_{\mathrm{exc},(j)}\right),
$$
    where $Y_{(j)}$ is the population total in cell $j$, corresponding to the defined sample total $y_{(j)}$ in (\ref{eqn:binary}); $y_{\mathrm{exc},(j)}$ is the total of the binary outcome variable for nonsampled units in cell $j$, with posterior predictive distribution
$$
    (y_{\mathrm{exc},(j)}|-) \sim \mathrm{Binomial} (N_j-n_j, \mbox{logit}^{-1}(\mu(x_j)).
$$
    As before, for finite-population inference we approximate $y_{\mathrm{exc},(j)}$ by its predictive mean.

    We collect the posterior samples of the $\bar{y}_{\mathrm{exc}, j}$'s for the continuous case or $y_{\mathrm{exc},(j)}$'s for the binary case to obtain the corresponding posterior samples of $\theta$ and then present the posterior summary for the estimate $\tilde{\theta}$.

    We compare our Bayesian nonparametric estimator with the classical estimator $\tilde{\theta}^{\,\mathrm{design}}$ in (\ref{ht}), which in theory is design-consistent (but which in practice can be compromised by the necessary choices involved in constructing classical weights that correct for enough factors without being too variable). We use the design-based variance estimator under sampling with replacement approximation in the R package {\tt survey} \citep{lumley10}, and the standard deviation is specified as
$$
    \textrm{sd}^{\,\mathrm{design}}=\frac{1}{n}\sqrt{\sum_{i=1}^n w_i^2(y_i-\tilde{\theta}^{\,\mathrm{design}})^2},
$$
where the weights $w_i$ have been renormalized to have an average value of 1.

\section{Simulation studies}
\label{simulation}
    We investigate the statistical properties of our method. First we apply a method proposed by \citet{cook06} for checking validity of the statistical procedure and the software used for model fitting. Next we implement a repeated sampling study given the population data to check the calibration of our Bayesian procedure.

\subsection{Computational coherence check}
\label{check}
    If we draw parameters from their prior distributions and simulate data from the sampling distribution given the drawn parameters, and then perform the Bayesian inference correctly, the resulting posterior inferences will be correct on average. For example, 95\%, 80\%, and 50\%  posterior intervals should contain the true parameter values with probability 0.95, 0.80, and 0.50. We will investigate the three coverage rates of population quantities and hyperparameters and check whether they follow a uniform distribution by simulation.

    Following \citet{cook06}, we repeatedly draw hyperparameters from the prior distributions, simulate population data conditional on the hyperparameters, sample data from the population, and then conduct the posterior inference. We choose $N =1,\!000,\!000$ and $n = 1, \!000$. For illustrative purposes, suppose there are $J^0 (=10)$ poststratification cells to ensure the sampled units will occupy all the population cells and the unit weights $w^0_{j}$ are assigned values from 1 to $J^0$ with equal probabilities. We set $N_j = N/J^0$ for $j=1,\dots, J^0$. In the later simulation study, we will investigate the performance in the scenario where several population poststratification cells are not occupied by the samples and ignored in the inference.

    After sampling the parameters $(\beta, \sigma, \tau, l, \delta)$ from their prior distributions described in (\ref{stan-prior-2}), where $\sigma \sim \textrm{Cauchy}^{+}(0, 2.5)$ for illustration, we generate realizations of the mean function $\vec{\mu}$ from $N(\vec{x}\beta, \Sigma)$, where the covariance kernel of $\Sigma$ is formed by the drawn $({\tau}, l, \delta)$. Then we generate the population responses, considering both continuous and binary cases. After generating the population data, we draw $n$ samples from the population with probability proportional to $1/w^0$. For the selected units, $i=1,\dots,n$, we count the unique values of weights defining the poststratification cells and the corresponding number of units in each cell, denoted as $w_{j}$ and $n_j$, for $j=1,\dots, J$, where $J$ is the number of unique values, \emph{i.e.}, the number of cells. We collect the survey respondents, $y_i$, for $i=1,\dots, n_j$, for each cell $j$.

    We use Stan for the posterior computation with Markov chain Monte Carlo (MCMC) algorithm. For the continuous cases, we keep the permuted 3,000 draws from 3 separate MCMC chains after 3,000 warm-up draws. For the binary cases, we keep 6,000 draws from 3 separate MCMC chains after 6,000 warm-up draws. In Stan, each chain takes around 1 second after compiling. Even though the sample size is large, we need only to invert a $J \times J$ matrix for the likelihood computations for the Gaussian process. This matrix inversion is the computational bottleneck in our method.

     \begin{figure}[t]
    \centering
        \includegraphics[scale=0.96]{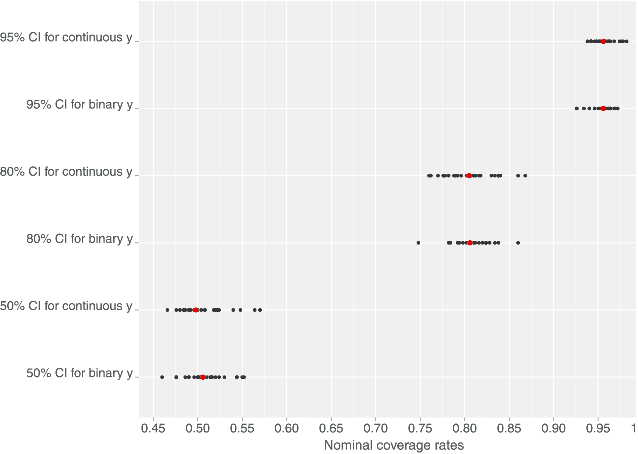}
    \caption{From the computational coherence check for the simulated-data example, coverage rates of central $95\%$, $80\%$, and $50\%$ posterior intervals for the population quantities and hyperparameters in the models for continuous and binary survey responses, correspondingly. Each dot represents a coverage rate. The red dots represent the median values.}
    \label{cr0817}
    \end{figure}

    First, we monitor the convergence of our MCMC algorithm  based on one data sample. The convergence measure we use is $\hat{R}$ based on split chains \citep{gelman13bda}, which when close to $1$ demonstrates that chains are well mixed. The collected posterior draws for our simulations have large enough effective sample sizes $n_\mathrm{eff}$ for the parameters, which is calculated conservatively by Stan using both cross-chain and within-chain estimates. The corresponding trace plots also show good mixing behaviors. The posterior median values are close to the true values. The posterior summaries for the parameters do not change significantly for different initial values of the underlying MCMC algorithm.

After checking convergence, we implement the process repeatedly 500
times to examine the coverage rates. We compute central $95\%$, $80\%$,
and $50\%$  intervals using the appropriate\,quantiles of the posterior
simulations.\,The coverage rates,\,shown in Figure \ref{cr0817}, are
close to their expected values for both the continuous and binary
survey responses.

    We also find the posterior inference for parameters to be robust under different prior distributions, such as inverse-Gamma distribution for $\tau^2$, Gamma distribution for $l^2$ or a heavy-tailed prior distribution for
    $\beta$.

\section{Simulation ranging from balanced weights to unbalanced weights}
\begin{figure}[t]
    \centering
        \includegraphics{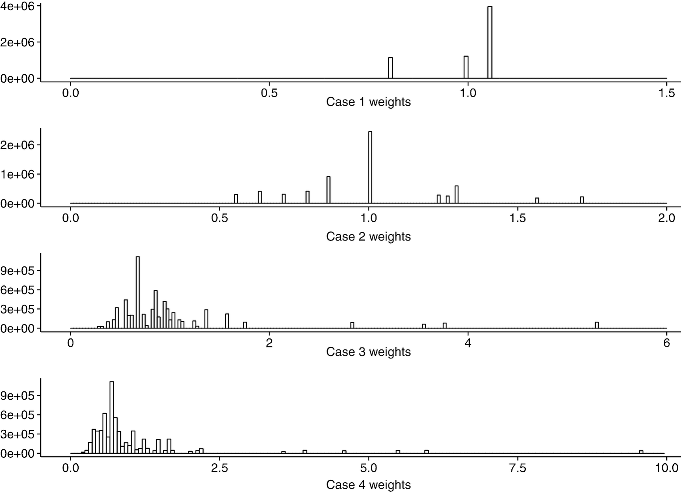}
    \caption{Frequency distributions of constructed unit weights in the population from four cases depending on different calibration variables: Case 1 adjusts for poverty gap; Case 2 adjusts for poverty gap and race; Case 3 adjusts for poverty gap, race and education; and Case 4 adjusts for poverty gap, race, education and gender. The four histograms are on different scales. }
    \label{wgh-freq}
    \end{figure}

We perform a simulation study comparing performances of the BNFP and the classical estimate under different sets of weights, ranging from calibration adjustment for only one factor to several factors and then getting successively less balanced, with more and more extreme weights. For constructing the weights in the population, we use external data sources to generate the population cell sizes and weights; this allows for cell counts and weights with values close to what is seen in practice. To this end, we borrow a cross-tabulation distribution from the 2013 New York City Longitudinal Survey of Well-being (LSW) conducted by the Columbia Population Research Center. The population in this case is adult residents of New York City, defined by the public cross-tabulations from the 2011 American Community Survey (ACS) with population size 6,307,213. Four different sets of weights are constructed via calibration depending on the choice of adjustment variables: 1) poverty gap (3 categories), as a poverty measure; 2) poverty gap and race (4 categories); 3) poverty gap, race and education (4 categories); and 4) poverty gap, race, education and gender. The cross-tabulations construct $J^0=3, 12, 48$ and $96$ cell weights $w^0$, respectively. The unit weights are obtained by the ratio factors $N_j^{\mathrm{ACS}}/n_j^{\mathrm{LSW}}$'s, where $N_j^{\mathrm{ACS}}$'s are the ACS cell counts and $n_j^{\mathrm{LSW}}$'s are the LSW sample cell sizes. We normalize the population weights to average to 1. The frequency distributions of the constructed weights are shown in Figure~\ref{wgh-freq}, which illustrates that the weights become successively less balanced. The last two cases have extreme weights, especially for Case 4, of which the maximum value is 9.60 and the minimum is 0.19.

Next we simulate the survey variable in the population. For the continuous case, a normal mixture model with different locations is used,
$$
Y \sim 0.3 \mathrm{N}(0.5w^2, \, 1) + 0.4 \mathrm{N}(5\log(w), \, 1) + 0.3 \mathrm{N}(5-w, \, 1).
$$
For the binary outcome,
$$
\mathrm{logit} \,\mbox{Pr}(Y=1) \sim \mathrm{N}(2 - 4(\log(w)+0.4)^2,\, 10).
$$
The motivation for the proposed models above is to simulate the population following a general distribution with irregular structure, such that we can investigate the flexibility of our proposed estimator. We perform 100 independent simulations, for each repetition drawing a sample from the finite population constructed above with units sampled with probabilities proportional to $1/w^0$. We set the sample size as 1000 for the continuous variable and as 200 for the binary
variable.

\begin{table}[t]
\begin{tabular}{lll rrrr}
 &Data&Estimator & Avg.\ S.E. & Bias & RMSE & Coverage$_{95}$ \\
  \hline
& Case 1 &BNFP & 0.07 & $-$0.00 & 0.06 & 97\% \\
 && classical & 0.06 & 0.01 & 0.06 & 97\% \\
 &Case 2&BNFP & 0.08 & $-$0.00 & 0.07 & 97\% \\
 Continuous&&classical & 0.07 & 0.01 & 0.07 & 96\% \\
($n=1000$)  & Case 3& BNFP & 0.13 & 0.03 & 0.13 & 95\% \\
&&  classical & 0.13 & $-$0.01 & 0.13 & 89\% \\
 & Case 4&BNFP & 0.17 & 0.09 & 0.30 & 85\% \\
&&  classical & 0.21 & $-$0.01 & 0.27 & 86\% \\
  \hline
&Case 1&BNFP& 0.03 & $-$0.01 & 0.04 & 93\% \\
 && classical & 0.04 & 0.00 & 0.04 &92\% \\
&Case 2&BNFP & 0.03 & $-$0.01 & 0.03 & 95\% \\
Binary && classical & 0.04 & 0.01& 0.04 & 95\% \\
$(n=200)$ &Case 3&BNFP & 0.03 & $-$0.00 & 0.03 & 99\% \\
&&classical & 0.04 & 0.00 & 0.04 & 96\% \\
&Case 4 & BNFP & 0.03 & 0.00 & 0.02  & 98\% \\
 && classical & 0.04 & $-$0.00 & 0.04  & 94\% \\
\end{tabular}
\caption{For simulated data from balanced weights to unbalanced weights, comparison between the Bayesian nonparametric finite population (BNFP) and classical design-based estimators on the average value of standard error (Avg S.E.), empirical bias, root mean squared error (RMSE), and coverage rates of $95\%$ intervals.}
\label{sim-3}
\end{table}

For each repetition, we also calculate the values of the classical
design-based estimator with its standard error. Table \ref{sim-3} shows
the general comparison:  BNFP estimation performs more efficiently and
consistently than the classical estimator in terms of smaller standard
error, smaller actual root mean squared error (RMSE) and better
coverage, while both of them are comparable for the reported bias. The
improved performances by BNFP estimation for binary outcomes are more
apparent than those with continuous outcomes mainly due to smaller
sample size. The improvement via BNFP estimation is more evident as the
weights become less balanced and much noisier. We note that for Case 4
with the continuous outcome, the coverages from both approaches are
relatively low and the bias and RMSE of BNFP are larger than the
classical estimator. We find that large values of bias and then large
RMSE of BNFP estimation correspond to samples during repetitions when
the maximum of the weights is selected but only one sample takes on
this value and the simulated $y$ is extremely large, which results in
an outlier. The smoothing property of GPs pulls the cell estimate
towards its neighbors. However, the partial pooling offered here is not
enough to eliminate the extreme influence of this outlier. This raises
the concern on the sensitivity of Bayesian modeling to highly variable
weights and calls for a more informative prior distribution. Another
concern is that the samples fail to collect all possible weights in the
population, especially extremely large weights, as an incompleteness of
our model, which treats the observed weights as the only possible cells
in the population. Consequently, when the outcome is strongly related
to the weights, the estimation could be biased. We will provide more
discussions on these issues in Section~\ref{conclusion}.

\begin{figure}[t!]
    \centering
    \includegraphics[scale=1.02]{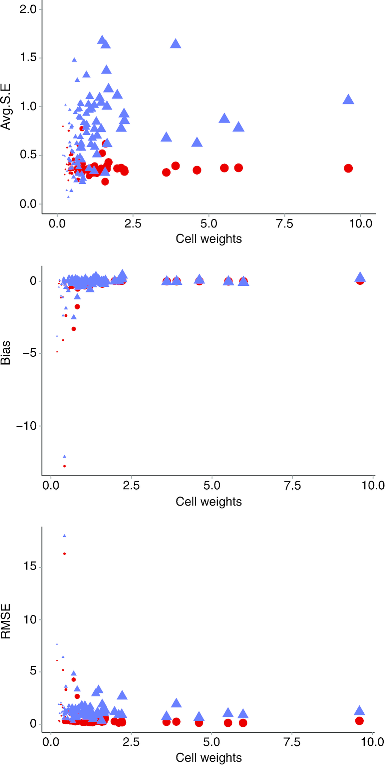}
     \caption{Comparison between the Bayesian nonparametric finite population (BNFP) and classical design-based estimators  on the average value of standard error (Avg S.E.), empirical bias, and root mean squared error (RMSE) on cell mean estimates. Each dot is one of the 96 population cells constructed by the unique weight values adjusting for poverty gap, race, education and gender in Case 4 with continuous outcome. The red circles represent BNFP estimation results and blue triangles represent classical results. }
    \label{sub-con-output}
    \end{figure}

We find that BNFP estimation outperforms the classical estimators in
small sample settings. Furthermore, we look at the subgroup estimates
for population cell means, which serve as intermediate outputs in the
BNFP estimation procedure. If extra information is available, the
subgroups of substantive interest can also be investigated. We compare
the poststratification cell mean estimates between BNFP and classical
estimates. The same conclusions apply for the four cases, which
supports that the BNFP estimation is more robust in particular for
small cell sizes with highly variable weights. To demonstrate the
advantage of BNFP estimation, we only present the results in Case 4 of
continuous outcome with the least balanced weights.
Figure~\ref{sub-con-output} shows that BNFP estimation generally yields
smaller standard error values and smaller RMSE comparing to the
classical estimates. The classical estimates for cell means are subject
to large variability and little efficiency. However, BNFP estimation
performs robustly. When no empty cells occur, BNFP estimation yields
good estimates and coverage rates for the population cell sizes $N_j$'s
and cell means $\mu_j$'s. In Case 4, empty cells always occur and the
observed weights cannot represent their true distribution. Our analysis
also shows that BNFP estimation is able to yield better coverage rates
with narrower credible intervals than the classical estimates, which is
not presented here.

Overall, the BNFP estimator performs better than classical estimates. When the sample sizes are large, the improvements are subtle. This fits our goals: as we are not incorporating substantive prior information, we would not expect our Bayesian approach to outperform classical methods when sample size is large relative to the questions being asked.  Rather, we wish to do as well as the classical approach with large samples, appropriately adjusting for the information in the weighting, and then perform better in settings involving partial pooling and small area estimation, thus gaining the benefits of Bayesian inference in a way that is consistent with sampling design.

\section{Application to Fragile Families and Child Wellbeing Study}
\label{application}

    The Fragile Families and Child Wellbeing Study \citep{ffdesign01} follows a cohort of nearly 5,000 children born in the U.S. between 1998 and 2000 and includes an oversample of nonmarital births. The study addresses the conditions and capabilities of unmarried parents, especially fathers, and the welfare of their children. The core study consists of interviews with both mothers and fathers at the child's birth and again when children are ages one, three, five, and nine. The sampling design is multistage (sequentially sampling cities, hospitals, and births) and complex--involving stratified sampling, cluster sampling, probability-proportional-to-size sampling, and random selection with predetermined quota. There are two main kinds of unit weights: national weights and city weights. Applying the national weights \citep{ffweights08} makes the data from the 16 randomly selected cities representative of births occurring in large cities in the U.S. between 1998 and 2000. The city weights make the records selected inside each city representative of births occurring in this city between 1998 and 2000 and adjust for selection bias, nonresponse rates, benchmarking calibration on mother's marital status, age, ethnicity, and education information at the baseline.

    For our example here, we work with the New York City weights and a binary survey response of whether the mother received income from public assistance or welfare or food stamp services. We perform two analyses, one for the baseline survey and one for the Year 9 follow-up survey. At baseline, the sample has $n=297$ births and $J=161$ poststratification cells corresponding to the unique values of the weights. Most of the nonempty cells in the sample are occupied by only one or two cases. In Year 9, the sample contained $n=193$ births, corresponding to $J=178$ unique values for the weights, with each cell containing 1 or 2 units. We implement the BNFP procedure using its default settings (as described in Section \ref{methods}) by Stan, running three parallel MCMC chains with 10,000 iterations, which we found more than sufficient for mixing of the chains and facilitates the posterior predictive checks as follows.

 \begin{table}
\begin{tabular}{ll rrr rr}
Data & Parameter & Mean & Median &SD& $2.5\%$ & $97.5\%$ \\
  \hline
Baseline & $\hat{p}$& 0.20 & 0.20 & 0.02 & 0.16 & 0.25\\
&$\beta$& $-$0.40 & $-$0.37 & 0.21 & $-$0.91 & $-$0.07 \\
&$\tau$& 1.74 & 1.53 & 0.92 & 0.69 & 4.10 \\
&$l$&1.75 & 1.01 & 17.69 & 0.30 & 3.92 \\
&$\delta$& 0.38 & 0.35 & 0.23 & 0.04 & 0.85 \\
    \hline
Year 9 &$\hat{p}$& 0.45 & 0.45 & 0.03 & 0.39 & 0.51 \\
&$\beta$& $-$0.34 & $-$0.26 & 0.39 & $-$1.36 & 0.21 \\
&$\tau$& 3.37 & 2.74 & 2.42 & 1.02 & 9.60 \\
&$l$& 1.44 & 1.32 & 0.70 & 0.40 & 3.13 \\
&$\delta$& 0.36 & 0.35 & 0.21 & 0.04 & 0.78
\end{tabular}
\caption{For the Fragile Families Study, Bayesian nonparametric finite-population (BNFP) inferences for model parameters for responses to the public support question. We find greater uncertainty and greater variability for the Year 9 follow-up, compared to the baseline.}
\label{ft}
\end{table}

    We assess the fit of the model to data using posterior predictive checks \citep{ppp:gelman-meng-stern96}. We generate replicated data $y^{rep}$ from their posterior predictive distributions given the model and the data and calculate the one sided posterior predictive $p$-values,
    $
    \mbox{Pr}(T(y^{rep}) \geq T(y)|y,\phi),
    $
where $T(y)$ is the test statistic and $\phi$ represents model parameters. A $p$-value close to 0 indicates a lack of fit in the direction of $T(y)$.  Here we use the sample total $y_{(j)}$ of responses in cells $j$ as the test statistics $T(y)$, following a general recommendation to check the fit of aspects of a model that are of applied interest, and we obtain $y_{(j)}^{rep,l}$ based on posterior samples of parameter $\phi^{l}$, for $l=1,\dots,L$. The posterior predictive $p$-value is computed as
$
\frac{1}{L}\sum_{l=1}^{L}I(y_{(j)}^{rep,l}\geq y_{(j)}),
$
for each cell $j=1,\dots,J$. At baseline, among the obtained $161$ posterior predictive $p$-values, the minimum is $0.70$. In Year 9, the $178$ posterior predictive $p$-values are all above $0.55$. Most $p$-values are around $0.75$, which does not provide significant evidence against the model, indicating the fitting performs well.

At baseline, BNFP estimation yields a posterior mean for the proportion getting public support $\hat{p}$ of
0.20 with standard error 0.023 and $95\%$ posterior interval $(0.16,
0.25)$. The BNFP estimator is larger than the design-based estimator of
0.16, which has a similar standard error of 0.024. In Year 9, the BNFP
posterior mean for $\hat{p}$ is 0.45 with standard error 0.030 and
$95\%$ posterior interval $(0.39, 0.51)$. The BNFP estimator is larger
than the design-based estimator of 0.32, which has a larger standard
error of 0.044. Comparing to the baseline survey when the child was
born, the estimate of the proportion of mothers receiving public
support services increases when the child was Year 9. This is expected
since child care in fragile families needs more social support as the
child grows up and some biological fathers become much less involved.

 \begin{figure}[h!]
 \centering
 \includegraphics{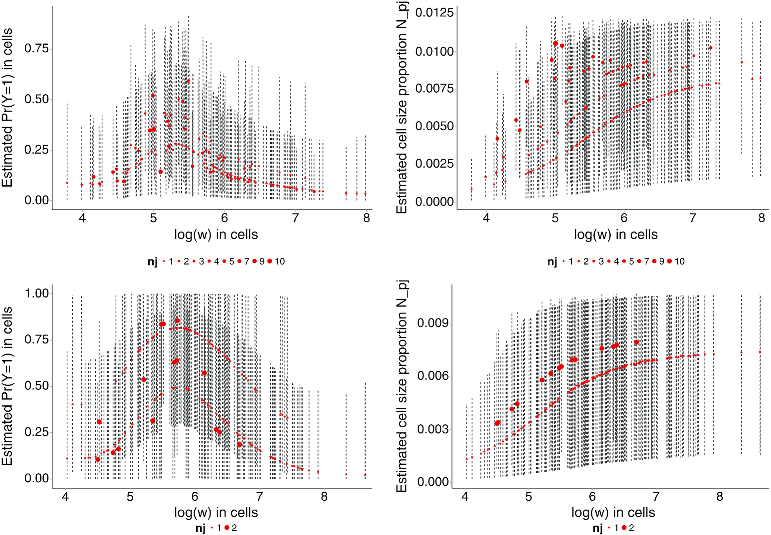}
 \caption{Estimated Fragile Families survey population poststratification cell mean probabilities $\theta_j$ and cell size proportions $N_{pj}$ for unit weights in different cells. Red dots are the posterior mean estimates and the black vertical lines are the 95\% posterior credible intervals. Dot sizes are proportional to sample cell sizes $n_j$; the top two plots are for the baseline survey and the bottom two are for the follow-up survey.}
 \label{ftnp}
\end{figure}

The posterior summaries for the hyperparameters are shown in Table \ref{ft}. The posterior mean of the scale $\tau$ is larger in the follow-up sample than in the baseline, which illustrates larger variability for the cell mean probabilities. As seen from Figure \ref{ftnp}, the poststratification cell mean probabilities depend on the unit weights in cells in an irregular structure. Because the $N_j$'s are not normalized, we look at the poststratification cell size proportions $N_{pj}(=N_j/\sum N_j)$. Since most sample cell sizes are 1, we see a general increasing pattern between the poststratification cell size proportions and the unit weights in cell $j$, for $j=1,\dots, J$. The uncertainty is large due to the small sample cell sizes. These conclusions hold for both the baseline and the follow-up survey. In the follow-up survey, the correlation between the cell mean probabilities is stronger and more variable compared to those in the baseline
survey.\looseness=-1

\section{Discussion}
\label{conclusion}

    We have demonstrated a nonparametric Bayesian estimation for population inference for the problem in which inverse-probability sampling weights on the observed units represent all the information that is known about the design. Our approach proceeds by estimating the weights for the nonsampled units in the population and simultaneously modeling the survey outcome given the weights as predictors using a robust Gaussian process regression model. This novel procedure captures uncertainty in the sampling as well as in the unobserved outcomes. We implement the model in Stan, a flexible Bayesian inference environment that can allow the model to be easily expanded to include additional predictors and levels of modeling, as well as alternative link functions and error distributions. In simulation studies, our fully Bayesian approach performs well compared to classical design-based inference. Partial pooling across small cells allows the Bayesian estimator to be more efficient in terms of smaller variance, smaller RMSE, and better coverage rates.

    Our problem framework is clean in the sense of using only externally-supplied survey weights, thus representing the start of an open-ended framework for model-based inference based on the same assumptions as the fundamental classical model of inverse-probability weighting. There are many interesting avenues for development. When the variables used to construct the weights are themselves available, they can be included in the model \citep{gelman07}. A natural next step is to model interactions among these variables. Multilevel regression and poststratification have achieved success for subpopulation estimations at much finer levels. More design information or covariates can be incorporated. Another direction to explore is to extend our framework when we have only partial census information, for example, certain margins and interactions but not the full cross-tabulations. Our approach can be used to make inference on the unknown census information, such as the population cell sizes $N_j$'s. This also facilitates small area estimation under a hierarchical modeling framework.

    Two concerns remain about our method. First, the inferential procedure assumes that all unique values of the weights have been observed. If it is deemed likely that there are some large weights in the population that have not occurred in the sample, this will need to be accounted for in the model, either as a distribution for unobserved weight values, or by directly modeling the factors that go into the construction of the weights. Under an appropriate hierarchical modeling framework, we can partially pool the data and make inference on the empty cells, borrowing information from observed nonempty cells. Second, in difficult problems (those with a few observations with very large weights), our inference will necessarily be sensitive to the smoothing done at the high end, in particular the parametric forms of the model for the $N_j$'s and the regression of $y|w$. This is the Bayesian counterpart to the instability of classical inferences when weights are highly variable. In practical settings in which there are a few observations with very large weights, these typically arise by multiplying several moderately-large factors. It should be possible to regain stability of estimation by directly modeling the factors that go into the weights with main effects and partially pooling the interactions; such a procedure should outperform simple inverse-probability weighting by virtue of using additional information not present in the survey weights alone. But that is a subject for future research. In the present paper we have presented a full Bayesian nonparametric inference using only the externally-supplied survey weights, with the understanding that inference can be improved by including additional information.

\bibliographystyle{ba}

\begin{acknowledgement}
We thank the National Science Foundation, the Institute of Education
Sciences, the Office of Naval Research, the Robin Hood Foundation, and
the Fragile Families Study project for partial support of this work.
\end{acknowledgement}





\end{document}